\documentclass[aps,prb,twocolumn,superscriptaddress]{revtex4}
\usepackage{epsf,graphicx}
\usepackage{amssymb}
\usepackage{amsmath}
\usepackage{latexsym,bm,array,amsfonts,multirow}
\usepackage{color}
\usepackage{ulem}
\begin{document}

\title{Superconducting fluctuations and charge-4$e$ plaquette state at strong coupling}
\author{Qiong Qin}
\affiliation{Beijing National Laboratory for Condensed Matter Physics and Institute of
	Physics, Chinese Academy of Sciences, Beijing 100190, China}
\affiliation{University of Chinese Academy of Sciences, Beijing 100049, China}
\author{Jian-Jun Dong}
\affiliation{Department of Physics and Chongqing Key Laboratory for Strongly Coupled Physics, Chongqing University, Chongqing 401331, China}
\author{Yutao Sheng}
\affiliation{Beijing National Laboratory for Condensed Matter Physics and Institute of
	Physics, Chinese Academy of Sciences, Beijing 100190, China}
\affiliation{University of Chinese Academy of Sciences, Beijing 100049, China}
\author{Dongchen Huang}
\affiliation{Beijing National Laboratory for Condensed Matter Physics and Institute of
Physics, Chinese Academy of Sciences, Beijing 100190, China}
\affiliation{University of Chinese Academy of Sciences, Beijing 100049, China}
\author{Yi-feng Yang}
\email[]{yifeng@iphy.ac.cn}
\affiliation{Beijing National Laboratory for Condensed Matter Physics and Institute of
Physics, Chinese Academy of Sciences, Beijing 100190, China}
\affiliation{University of Chinese Academy of Sciences, Beijing 100049, China}
\affiliation{Songshan Lake Materials Laboratory, Dongguan, Guangdong 523808, China}
\date{\today}

\begin{abstract}
We apply the static auxiliary field Monte Carlo approach to study  phase correlations of the pairing fields in a model with spin-singlet pairing interaction. We find that the short- and long-distance phase correlations are well captured by the phase mutual information, which allows us to construct a theoretical phase diagram containing the uniform $d$-wave superconducting region, the phase fluctuating region, the local pairing region, and the disordered region. We show that the gradual development of phase coherence has a number of consequences on spectroscopic measurements, such as the development of the Fermi arc and the anisotropy in the angle-resolved spectra, scattering rate, entropy, specific heat, and quasiparticle dispersion, in good agreement with experimental observations. For strong coupling, our Monte Carlo simulation reveals an unexpected charge-4$e$ plaquette state with $d$-wave bonds, which competes with the uniform $d$-wave superconductivity and exhibits a U-shaped density of states.  
\end{abstract}
\maketitle

\section{Introduction}
Superconducting fluctuations have been proposed to play an important role in underdoped cuprates \cite{Emery1995,Franz1998,Kwon2001,Norman2007,Samokhin2004,Mayr2005a,Eckl2002,Han2010,Zhong2011,Berg2007,Tesanovic2008,Banerjee2011,Allais2014,Singh2021}. Their presence may be responsible for the back bending bands above the superconducting transition temperature $T_c$ \cite{Kanigel2008}, continuous variation of the spectral gap across the transition \cite{Ding1996}, and probably the large Nernst effect and diamagnetic signals \cite{Li2010}. They have also been used to explain the mysterious pseduogap state \cite{Gomes2007,Yang2008a,Lee2009,Zhou2019,Bilbro2011,Civelli2005,Macridin2006,Greco2009,Ferrero2010,Sordi2012,Gull2013,Wu2018,Richie-Halford2020,Jiang2022,Long2023,Kyung2004,Kyung2006,Wu2017a,Vucicevic2017}, but negated by some experiments showing that superconducting fluctuations only exist in a much narrower region than the pseudogap \cite{Kondo2011}. Their interplay with competing orders may be the cause of particle-hole asymmetry \cite{He2011}, time-reversal symmetry breaking \cite{Kaminski2002,He2011}, or rotational symmetry breaking \cite{Gupta2021} observed in some materials.

In overdoped cuprates, mean-field analyses have been widely used to describe superconductivity, since it correctly predicted the $d$-wave pairing and the decrease in $T_c$ with hole density \cite{Emery1995}, while superconducting fluctuations have scarcely been considered seriously \cite{Rourke2011}, although experiments have reported a linear relation between the superfluid density and $T_c$ and thus highlighted the crucial role of superconducting phase stiffness \cite{Bozovic2016,Mahmood2019}. Very recently, the angle-resolved photoemission spectroscopy (ARPES) observation of a $d$-wave gap and particle-hole symmetric dispersion above $T_c$ in overdoped Bi$_2$Sr$_2$CaCu$_2$O$_{8+\delta}$ \cite{He2021,Chen2022,Zou2022} has stimulated intensive debates concerning the existence of phase fluctuations in overdoped cuprates and whether the observed anomalous properties are due to superconducting fluctuations or involve other mechanisms such as anisotropic impurity scattering \cite{Wang2022}.

In this work, we explore the potential consequences of superconducting fluctuations on the spectroscopic observations in overdoped cuprates. Different from previous studies \cite{Li2021a,Wang2022,Singh2021,Wei2022,Lee-Hone2018}, we employ a static auxiliary field Monte Carlo approach \cite{Dong2021a,Mukherjee2014,Liang2013,Dubi2007,Pasrija2016,Karmakar2020} and use phase mutual information to analyze short- and long-range phase correlations of the superconducting pairing fields. The mutual information \cite{Cover2006,Kraskov2004PRE,Varanasi1999,Gao2015,Khan2007,Belghazi2018PMLR,Poolel2019PMLR} measures the nonlinear association of the probabilistic distribution \cite{Speed2012,Reshef2011,Kinney2014} and has been sucessfully applied to various physical systems \cite{Koch-Janusz2018,Nir2020,Gokmen2021,ParisenToldin2019,Walsh2021a,Nicoletti2021}. It provides an excellent indicator of superconducting phase correlation and allows us to construct a superconducting phase diagram with the temperature and pairing interaction. We identify three temperature scales over a wide intermediate range of the pairing interaction, and we determine four distinct phases: the superconducting, (macroscopic) phase fluctuating, local pairing, and disordered regions. Calculations of the angle-resolved spectra, scattering rate, entropy, specific heat, quasiparticle dispersion, and Fermi arc show interesting anisotropic features, beyond the mean-field theory but agreeing well with experiments. For sufficiently strong pairing interaction, we find a plaquette state of charge-4$e$ pairing with a U-shaped density of states that competes with the uniform $d$-wave superconductivity \cite{Kim2022}. Our work provides a systematic understanding of the effects of superconducting fluctuations on the spectroscopic properties in overdoped cuprates.

\section{Model and method}
We start with the following Hamiltonian,
\begin{eqnarray}
H=-\sum_{il\sigma}t_{il}c_{i\sigma}^{\dagger}c_{l\sigma}-\mu\sum_{i\sigma}c_{i\sigma}^{\dagger}c_{i\sigma}-V\sum_{\langle ij\rangle}\left(\psi_{ij}^{\rm S}\right)^{\dagger}\psi_{ij}^{\rm S},
\label{model}
\end{eqnarray}
where the pairing interaction is written in an explicit form for the spin-singlet superconductivity with $\psi_{ij}^{\rm S}=\frac{1}{\sqrt{2}}(c_{i\downarrow}c_{j\uparrow}-c_{i\uparrow}c_{j\downarrow})$ and the strength $V>0$, which may be directly derived from an antiferromagnetic spin-density interaction or an attractive charge-density interaction between nearest-neighbor sites \cite{Monthoux2007}. To decouple the pairing interaction, we apply the Hubbard-Stratonovich transformation and introduce the auxiliary field $\Delta_{ij}$ \cite{Coleman2015}: 
\begin{equation}
	-V\bar{\psi}_{ij}^{\rm S}\psi_{ij}^{\rm S}\rightarrow \sqrt{2}\left(\bar{\Delta}_{ij}\psi_{ij}^{\rm S}+\bar{\psi}_{ij}^{\rm S}\Delta_{ij}\right)+\frac{2|\Delta_{ij}|^2}{V}.
	\end{equation}
The model is generally unsolvable. To proceed, we further adopt a static approximation and ignore the imaginary time dependence of the auxiliary fields. This allows us to integrate out the fermionic degrees of freedom and simulate solely the pairing fields $\Delta_{ij}$. We obtain an effective action:
\begin{equation}
S_{\rm eff}(\Delta)=-\sum_{ i}\ln(1+e^{-\beta\Lambda_i})+\frac{2\beta}{V}\sum_{\langle ij\rangle}|\Delta_{ij}|^2,
\end{equation}
where $\beta$ is the inverse temperature and $\Lambda_i$ are the eigenvalues of the matrix
\begin{eqnarray}
O=\left(\begin{array}{cc}
-\mu-T&M\\
M^{*}&\mu+T\\
\end{array}\right),
\end{eqnarray}
in which $T$ is the $N\times N$ hopping matrix ($N$ is the site number) and $M_{ij}=\Delta_{ij}$ comes from the pairing term. 

\begin{figure}[ptb]
	\begin{center}
		\includegraphics[width=8.6cm]{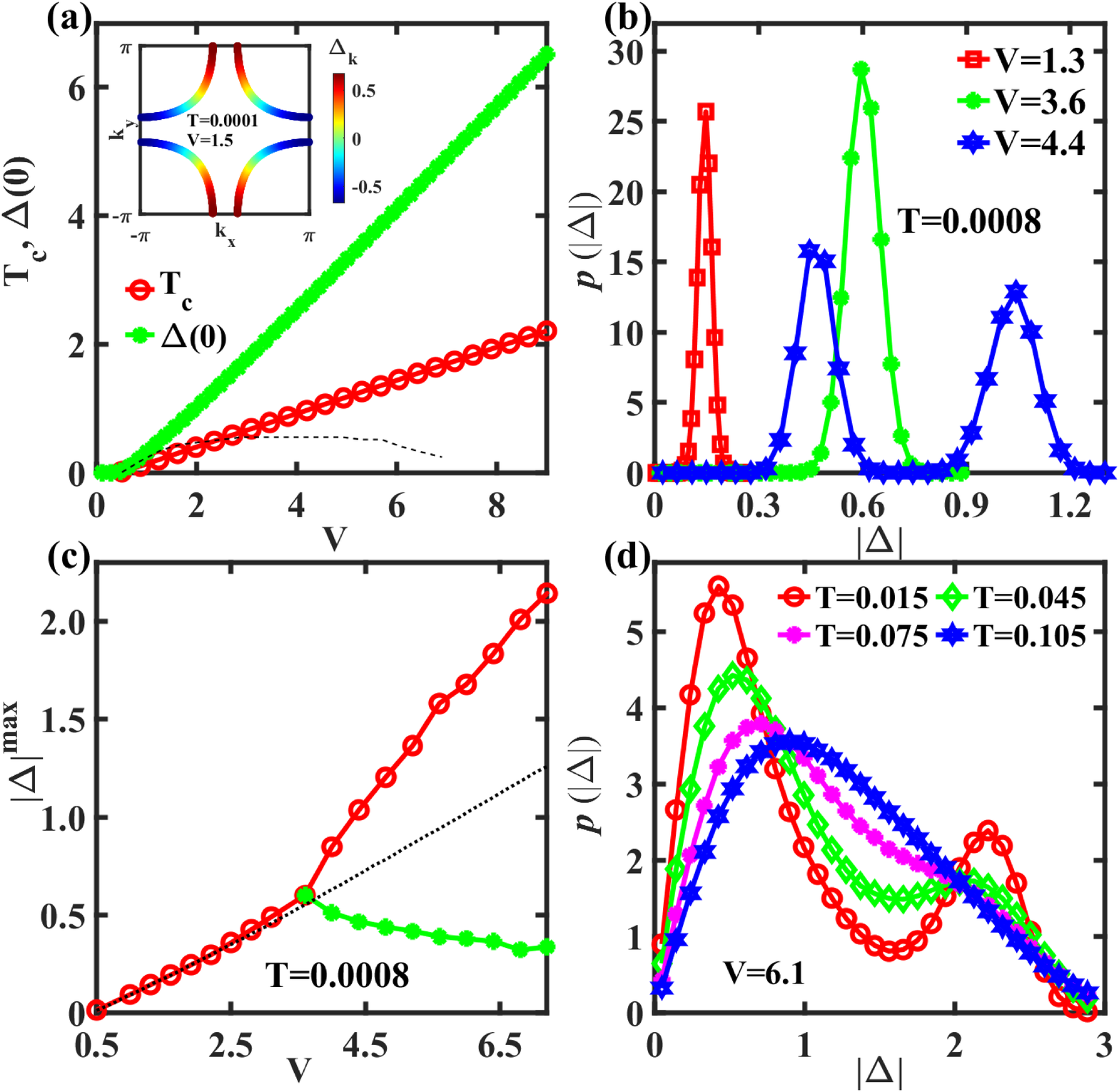}
	\end{center}
	\caption{(a) The mean-field phase diagram, where $T_c$ and $\Delta(0)$ are the superconducting transition temperature and the maximum of the momentum-dependent gap at $T=0$, respectively. For comparison, the dashed line gives the values of $8T_c$ from our static auxiliary field Monte Carlo simulations to be discussed later. The inset gives the superconducting gap $\Delta_{\bm{k}}$ along the Fermi surface for $V=1.5$ at $T=0.0001$. (b) Evolution of the amplitude distribution $p( |\Delta|)$ for all bonds at $T=0.0008$, showing one peak for moderate interaction and two peaks for strong interaction. (c) The peak position $|\Delta|^{\rm max}$ of $p( |\Delta| )$ as a function of the pairing  interaction $V$, where two maxima are seen to occur for $V\ge 4$. The dotted line shows the uniform mean-field solution for comparison. (d) Evolution of $p( |\Delta| )$ from two-peak to one-peak structure with increasing temperature for $V=6.1$.}
	\label{fig1}
\end{figure}

For spin-singlet pairing, $\Delta_{ij}$ is symmetric and defined on the bond between two nearest-neighbor sites $ij$. We thus have totally $2N$ independent complex variables satisfying the probabilistic distribution: 
\begin{equation}
p(\Delta)=Z^{-1}e^{-S_{\rm eff}},\ \ \ \ Z=\int \mathcal{D}\Delta\mathcal{D}\bar{\Delta} e^{-S_{\rm eff}},
\end{equation}
where $Z$ is the partition function serving as the normalization factor. Because $O$ is an Hermitian matrix, all its eigenvalues $\Lambda_i$ and consequently $S_{\rm eff}$ are real. Hence, the above model can be simulated using Monte Carlo with the Metropolis algorithm. In the following, all presented results are obtained on a $10\times10$ square lattice ($N=100$). We have also performed calculations on a $12\times 12$ lattice, and the results are qualitatively consistent. Unfortunately, due to computational limitations, we cannot go to a larger size and conduct a comprehensive finite-size scaling analysis to obtain more accurate transition temperatures. But the qualitative physics is insensitive to the lattice size,  and our method is also verified for the standard XY model. For simplicity, only the nearest-neighbor ($t$) and next-nearest-neighbor ($t^\prime$) hopping parameters are included. We take $t^\prime=-0.45 t$ following the common choice in the literature \cite{Carbotte1994,Monthoux2001} and set $t$ as the energy unit. For real materials, $t$ is typically of the order of 0.1 eV. The chemical potential $\mu$ is specially chosen to be -1.4. As plotted in the inset of Fig.~\ref{fig1}(a), the corresponding noninteracting dispersion always gives a large Fermi surface as in overdoped cuprates \cite{Plate2005,Vignolle2008}.

\section{Results}
For comparison, we first discuss the uniform mean-field solution. The pairing fields are found to satisfy $\Delta^x=-\Delta^y$, where the superscript represents the bond direction. A gap along the Fermi surface is shown in the inset of Fig.~\ref{fig1}(a), reflecting a typical $d_{x^2-y^2}$-wave structure \cite{Monthoux1991}.  The maximum gap size $\Delta(T=0)$ and $T_c$ are plotted in Fig.~\ref{fig1}(a) and both increase with increasing pairing interaction $V$. The typical BCS formula of $T_c$ is reproduced only at small $V$ but violated for $V>0.5$, where we find a roughly linear relation $T_c\sim V$ with the ratio $2\Delta(0)/T_c\approx 4.6-6.1$, which differs from the predictions of the weak-coupling BCS theory. A significant reduction of $T_c$ (dashed line) is found once superconducting fluctuations are included.  

\subsection{Spatial correlations of the pairing fields}
Our Monte Carlo simulations of the auxiliary pairing fields allow us to study the effect of superconducting fluctuations beyond the mean-field solution. Figure~\ref{fig1}(b) shows the amplitude distribution of the pairing field $p(|\Delta|)$ on all bonds at a very low temperature $T=0.0008$. We focus on moderate and large pairing interactions where $T_c$ is not too small for our numerical simulations. For $V=1.3$ and 3.6, the distributions are quite normal and can be well fitted by a Gaussian form. But for $V\ge 4$, it develops a two-peak structure. Figure~\ref{fig1}(c) summarizes the peak positions as a function of $V$ for $T=0.0008$. Compared to the uniform mean-field solution (dotted line), a transition occurs at $V\approx 4.0$, separating the superconductivity into two regions. We will see that they correspond to a homogeneous superconducting state for moderate $V$ and a spatially modulated state for large $V$, respectively. In Fig.~\ref{fig1}(d), the two-peak distribution at large $V$ is gradually suppressed with increasing temperature and becomes a single peak at sufficiently high temperatures. Apart from this, however, the amplitude distribution seems to lack distinctive features in its temperature evolution. We therefore explore mainly the phase fluctuations in the following sections.

\begin{figure}[t]
	\begin{center}
		\includegraphics[width=8.6cm]{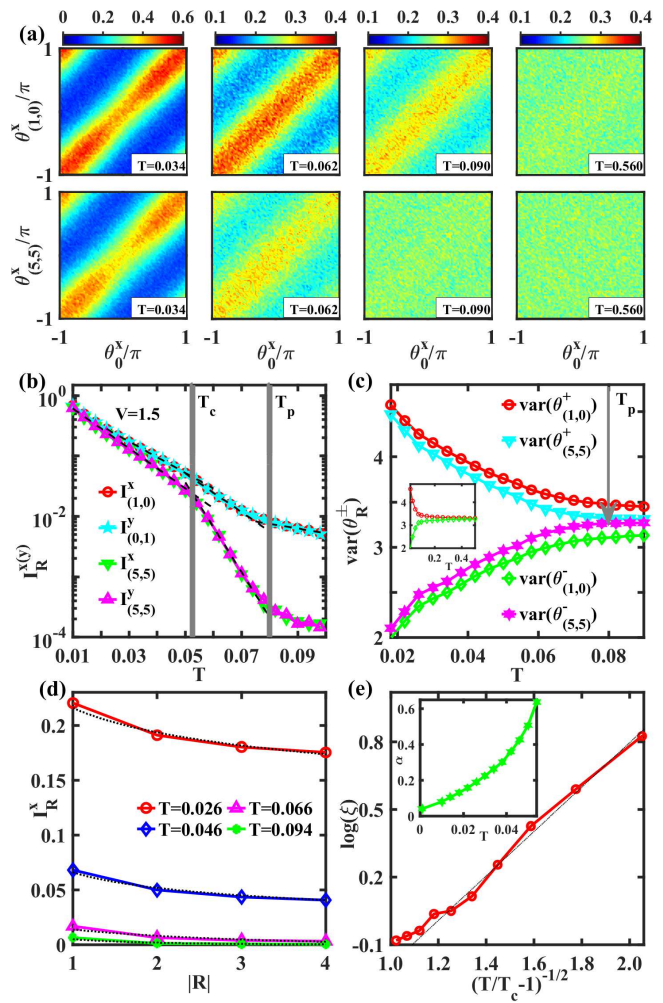}
	\end{center}
\caption{(a) Comparison of the joint distribution $p(\theta^x_{\bm{0}},\theta^x_{\bm{R}})$ for $\bm{R}=(1,0)$ and $\bm{R}=(5,5)$ at different temperatures. (b) Evolution of the short- and long-distance phase mutual information calculated from (a) as a function of temperature, showing two temperature scales $T_c$ and $T_p$ (vertical grey lines) from the slope change. (c) Temperature dependence of the variance of two principal components $\theta^{\pm}_{\bm{R}}=\frac{1}{\sqrt{2}}(\theta_{\bm{0}}\pm \theta_{\bm{R}})$ from PCA analyses of the data in (a) for short- and long-distance phase correlations. The inset shows the results for $\bm{R}=(1,0)$ on a larger temperature window. (d) Decay of the phase mutual information $I^x_{\bm{R}}$ with distance $|\bm{R}|=|R_x|+|R_y|$. The dotted lines give the power-law fit using $I^x_{\bm{R}}\propto |\bm{R}|^{-\alpha}$ below $T_c$ and the exponential fit $I^x_{\bm{R}}\propto \exp(-|\bm{R}|/\xi)|$ above $T_c$, respectively. (e) The extracted correlation length $\xi$ versus $(T/T_c-1)^{-1/2}$ following the BKT prediction (dashed line) for $T>T_c$. The inset shows the extracted exponent $\alpha$ for $T<T_c$.}
	\label{fig2}
\end{figure}

We first focus on the homogeneous state for moderate $V$ and study its properties from the perspective of phase correlations of the pairing fields. Our tool is the joint distribution $p(\theta_{\bm{0}}^{i},\theta_{\bm{R}}^i)$, where $\bm{0}\equiv(0,0)$ denotes the bond attached to any origin site, $\bm{R}$ represents the relative coordinate of the other bond, and $i=x,y$ denotes the bond along the $x$- or $y$-direction. Figure~\ref{fig2}(a) plots some typical results for $i=x$ and $\bm{R}=(1,0)$ (short-distance) and $(5,5)$ (long-distance) at different temperatures. Due to rotational symmetries, the results are the same for $i=y$. At high temperatures, we find a uniform distribution due to strong thermal fluctuations. With lowering temperature, two phases are gradually locked, as manifested by the maximum distribution along the diagonal. A direct comparison shows that this feature first appears on short range with $\bm{R}=(1,0)$ and then on longe range with $\bm{R}=(5,5)$. Hence, the phase coherence of the superconducting pairing grows gradually on the lattice to longer distance with decreasing temperature.

To quantify the correlation, we introduce their phase mutual information defined as 
 \begin{eqnarray}
 I^i_{\bm{R}}=\int d\theta^i_{\bm{0}} d\theta^i_{\bm{R}}~ p(\theta^i_{\bm{0}},\theta^i_{\bm{R}})\ln\frac{p(\theta^i_{\bm{0}},\theta^i_{\bm{R}})}{p(\theta^i_{\bm{0}})p(\theta^i_{\bm{R}})},
 \end{eqnarray}
where $p(x)$ is the marginal distribution function of the continuous random variable $x$ and $p(x,y)$ is the joint probability distribution of $x$ and $y$. Figure~\ref{fig2}(b) compares the  phase correlations as a function of temperature on short and long distances. We see they all exhibit similar behavior below $T_c=0.054$ and vary exponentially (dashed lines) with the temperature. But for $\bm{R}=(5,5)$, the mutual information suffers from an abrupt change in its temperature dependence and diminishes more rapidly above $T_c$. Such a slope change actually occurs in the phase mutual information at all distances, but is responsible for the deviation of the phase mutual information at different but close $|R|$ due to its more rapidly decay with distance above $T_c$. Thus, $T_c$ marks a characteristic temperature scale separating the phase coherence on different spatial scales, above which long-range correlations are more rapidly suppressed.

At higher temperature $T_p=0.08$ for the chosen parameters, a weaker slope change is found for both short- and long-distance correlations. To see what happens at this temperature, we apply the principal component analysis (PCA) to the Monte Carlo samples as collected in Fig.~\ref{fig2}(a). As expected, this reveals two principal directions $\theta^{\pm}_{\bm{R}}=\frac{1}{\sqrt{2}}(\theta_{\bm{0}}\pm \theta_{\bm{R}})$ on the $(\theta_{\bm{0}},\theta_{\bm{R}})$ plane for all temperatures, with opposite temperature dependence of their variances. The superscript $i$ is dropped because the data on both bond directions $i=x,y$ are considered together. As shown in Fig.~\ref{fig2}(c), the decrease of ${\rm var}(\theta^-_{\bm{R}})$ signifies the increase of phase locking degree on the distance $\bm{R}$ with lowering temperature. Interestingly, ${\rm var}(\theta^{\pm}_{\bm{R}})$ become almost equal above $T_p$ along both directions for $\bm{R}=(5,5)$, implying a uniform distribution on the $(\theta_{\bm{0}},\theta_{\bm{R}})$ plane and hence the almost complete loss of phase correlation on long distances. On the other hand, the two variances still differ for $\bm{R}=(1,0)$, indicating the existence of short-range correlation. The latter is to be suppressed only at much higher temperatures above $T_l=0.25$, as shown in the inset of Fig.~\ref{fig2}(c). Thus, $T_l$ marks a temperature scale above which no phase correlations are present (a disordered state). Below $T_l$, the pairing fields start to develop phase correlations between neighboring bonds, indicating the onset of local pairing only. Phase correlations on longer distances only emerge below $T_p$ in the phase fluctuating state and eventually grow into a quasi-long-range order [two-dimensional (2D) superconductivity] at $T_c$. 

The above separation of different regions may be seen from a different angle by plotting the mutual information as a function of the ``distance" $|\bm{R}|\equiv|R_x|+|R_y|$ (not the Euclidean distance). The results of $I^x_{\bm{R}}$ are shown in Fig.~\ref{fig2}(d). We observe power-law decay below $T_c$, characterized by $I^x_\textbf{R}\propto |\bm{R}|^{-\alpha}$, while above $T_c$, the data can be fitted with an exponential function, $I^x_\textbf{R}\propto \exp(-|\bm{R}|/\xi)$. The data saturate for larger distances below $T_c$, reflecting the quasi-long-range order under finite lattice size. We have examined these different behaviors in the standard XY model with the famous Berezinskii-Kosterlitz-Thouless (BKT) transition \cite{Berezinskii1972, Kosterlitz1973, Kosterlitz1974}. A slight difference is that, the finite size effect seems less pronounced in the XY model, possibly due to the inclusion of long-range interactions beyond nearest neighbors by integrating out the fermions in our model. Although finite-size effects cut off the power-law decay with distance below $T_c$, the temperature dependence of the phase mutual information still effectively captures the transition point for both models. Interestingly, the extracted correlation length $\xi$ also follows closely the scaling, $\log(\xi)\propto (T/T_c-1)^{-1/2}$, predicted by the BKT transition, and deviates above the phase fluctuation transition temperature $T_p$. Below $T_c$, the decay rate $\alpha$ extracted from the power-law scaling of $I^x_{\bm{R}}$ decreases with temperature. As shown in the inset, it remains a small value and approaches zero as $T\rightarrow 0$, indicating the establishment of true long-range order in the limit of zero temperature.

\begin{figure}[ptb]
	\begin{center}
		\includegraphics[width=8.6cm]{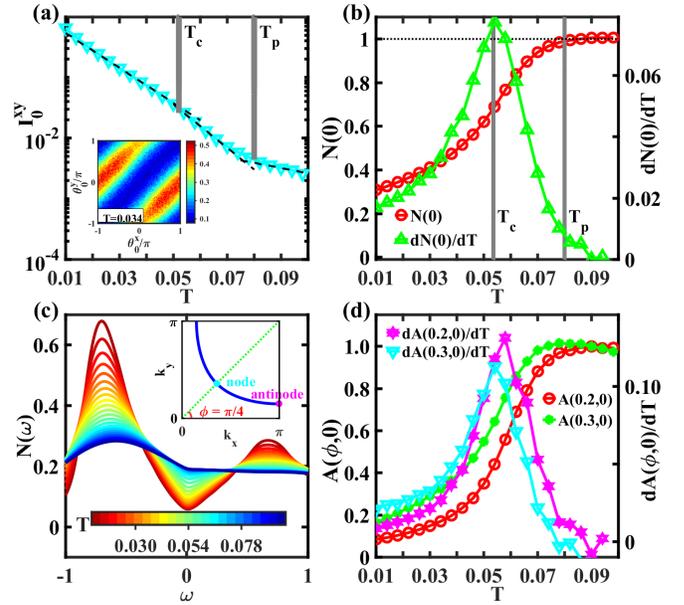}
	\end{center}
	\caption{(a) Temperature dependence of the phase mutual information $I^{xy}_{\bm{0}}$ between the $x$- and $y$-bonds attached to the same site. The inset shows their joint phase distribution at $T=0.034$, indicating $d$-wave correlations between two bonds. (b) The normalized total density of states $N(0)$ at the Fermi energy and its temperature derivative $dN(0)/dT$ as functions of temperature, showing features at $T_c$ and $T_p$ (grey vertical lines) determined from the phase mutual information. (c) Temperature evolution of the total density of states $N(\omega)$, showing the gradual gap opening near the Fermi energy. The inset illustrates the azimuthal angle $\phi$ and the positions of node and antinode. (d) Temperature dependence of the angle-resolved spectral function $A(\phi,0)$ and its derivative $dA(\phi,0)/dT$ at the noninteracting Fermi wave vector and the Fermi energy at $V=1.5$ for $\phi=0.2$, $0.3$.}
	\label{fig3}
\end{figure}

\subsection{Effects on spectroscopic properties}
Having established how the superconductivity is developed from its phase correlation, we now examine how these may be related to the experimental observations in real materials. First of all, the $d$-wave nature of the superconducting pairing can be seen from the joint distribution of $\theta^x_{\bm{0}}$ and $\theta^y_{\bm{0}}$ connected to the same site. As shown in the inset of Fig.~\ref{fig3}(a), we find a rough correlation, $\theta^y_{\bm{0}}=\theta^x_{\bm{0}} \pm \pi$, namely a sign change of the pairing fields along two perpendicular bond directions. Here we define the mutual information between $\theta^x_{\bm{0}}$ and $\theta^y_{\bm{0}}$ attached to the same site along two perpendicular directions:
	\begin{eqnarray}
		I^{xy}_{\bm{0}}=\int d\theta^x_{\bm{0}} d\theta^y_{\bm{0}}~ p(\theta^x_{\bm{0}},\theta^y_{\bm{0}})\ln\frac{p(\theta^x_{\bm{0}},\theta^y_{\bm{0}})}{p(\theta^x_{\bm{0}})p(\theta^y_{\bm{0}})}.
	\end{eqnarray}  
As we can see in Fig.~\ref{fig3}(a), its temperature dependence exhibits similar slope changes at $T_c$ and $T_p$. 

The separation of phase correlations on short and long distances has important consequences on the spectral properties, which may be studied by assuming a twist boundary condition to overcome the finite-size effect \cite{Li2018PRL}. Figure~\ref{fig3}(b) plots the total density of states at the Fermi energy $N(0)$ normalized by its high-temperature value. It is almost a constant above $T_p$, but then decreases gradually with lowering temperature, reflecting the spectral weight depression induced by a gap opening at zero energy. Interestingly, its temperature derivative $dN(0)/dT$ exhibits a maximum at around $T_c$, consistent with the slope change of the phase mutual information $I_\textbf{(5,5)}^{x(y)}$. Correspondingly, as shown in Fig.~\ref{fig3}(c), a pseudogap develops gradually on $N(\omega)$ with lowering temperature over the intermediate range $T_c<T<T_p$.  These establish a close relation between the phase correlation and the spectral gap of the superconductivity.

A similar temperature evolution is also seen in the angle-resolved spectral function $A(\phi,0)$ and its temperature derivative $dA(\phi,0)/dT$ along the azimuthal angle $\phi$ at the noninteracting Fermi wave vector and zero energy. For larger $\phi$ away from the antinode, the spectral function grows to a maximum at lower temperature and has a higher residual value at the zero-temperature limit. Meanwhile, its temperature derivative becomes more enhanced below $T_c$ but suppressed above $T_c$. Such an intrinsic anisotropy has been observed in the latest ARPES experiment \cite{Chen2022}.

 \begin{figure}[ptb]
 	\begin{center}
 		\includegraphics[width=8.6cm]{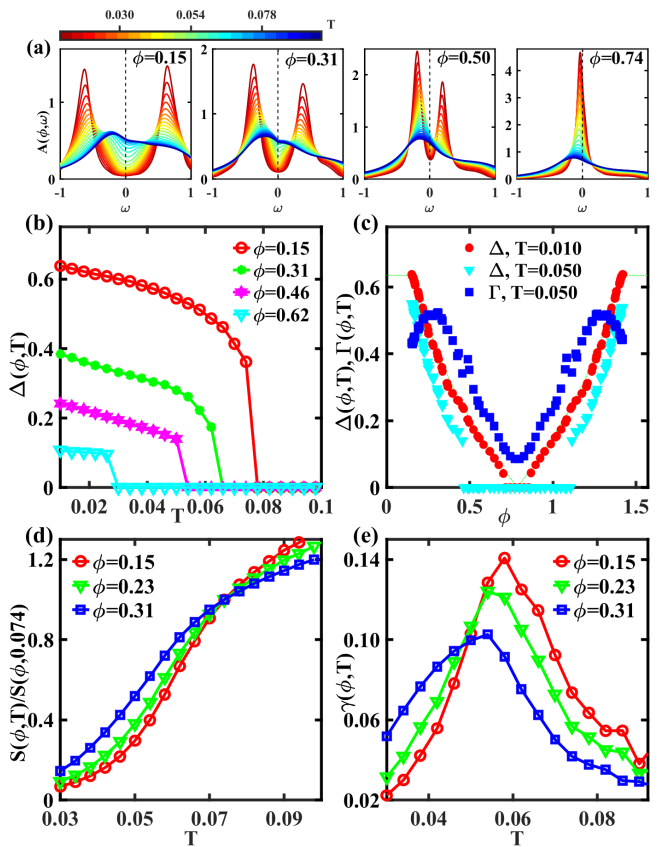}
 	\end{center}
 	\caption{(a) Temperature evolution of the angle-resolved spectral function $A(\phi,\omega)$ on different positions of the Fermi surface. (b) Comparison of the extracted gap $\Delta(\phi,T)$ from (a) as functions of the temperature $T$. (c) Angular dependence of the spectral gap $\Delta(\phi,T)$ and scattering rate $\Gamma(\phi,T)$ on the Fermi surface. $\Delta$ and $\Gamma$ are defined as the energy and the half-maximum half-width of the upper peak of the spectral function $A(\phi,\omega)$. (d) and (e) give the calculated thermal entropy $S(\phi,T)$ and specific heat coefficient $\gamma(\phi,T)$ as functions of temperature at different positions ($\phi$) on the Fermi surface.}
 	\label{fig4}
 \end{figure}

To clarify the origin of the anisotropy, we compare in Fig.~\ref{fig4}(a) the temperature dependence of the spectral functions $A(\phi,\omega)$ for different azimuthal angle $\phi$. Obviously, they exhibit very different behaviors near nodal or antinodal directions. Figure~\ref{fig4}(b) plots the extracted spectral gap $\Delta(\phi, T)$ as a function of temperature. With increasing temperature, the gap closes first near the nodal direction. Thus as shown in Fig.~\ref{fig4}(c), it only satisfies the ideal $d$-wave form $\Delta(\phi, T)\propto \cos(2\phi)$ (green dashed line) at sufficiently low temperatures. This is beyond the mean-field approximation but reflects the effect of phase fluctuations. Consequently, the scattering rate $\Gamma(\phi)$ estimated from the half-maximum half-width of the upper peak of $A(\phi,\omega)$ also exhibits smaller values near the node.

The anisotropy of the spectral functions has an effect on the angle-resolved thermal entropy $S(\phi,T)$ and the specific heat coefficient $\gamma(\phi,T)=dS(\phi,T)/dT$ by using $S(\phi,T)=-\int d\omega A(\phi,\omega)[f\ln f+(1-f)\ln(1-f)]$, where $f$ is the Fermi distribution function. As shown in Figs.~\ref{fig4}(d) and \ref{fig4}(e), the resulting $S(\phi,T)$ and $\gamma(\phi,T)$ exhibit similar temperature and angle dependence to $A(\phi,0)$ and $dA(\phi,0)/dT$ in Fig.~\ref{fig3}(d), which agree well with the entropy reduction and specific-heat anisotropy reported in the latest ARPES experiment \cite{Chen2022}.

\begin{figure}[ptb]
 	\begin{center}
 		\includegraphics[width=8.6cm]{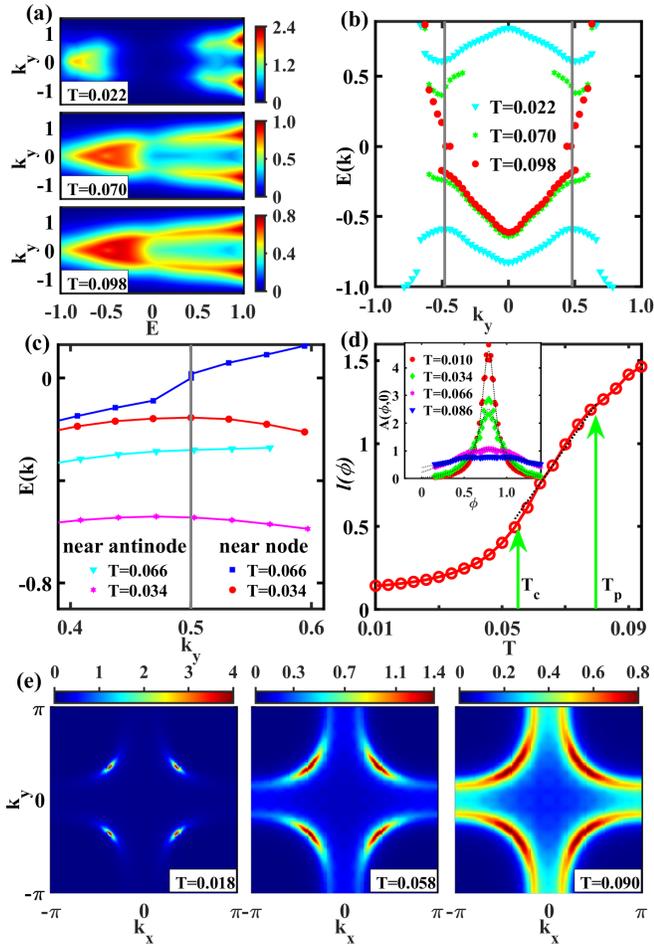}
 	\end{center}
 	\caption{ (a) Intensity plot of the spectral function $A(k_y,\omega)$ for $k_x=-3.047$ at $T=0.022$, 0.070, 0.098. (b) Extracted dispersions from the spectral functions at different temperatures, showing back bending even above $T_c$. The vertical grey lines mark the Fermi wave vector $k_y=\pm 0.4712$. (c) The dispersions near antinode and node for $T=0.066$ and 0.034. For comparison, all curves are shifted such that the Fermi wave vectors are located at $k_y=0.5$ (grey line). For clarity, only the lower (negative energy) parts of the superconducting dispersions are shown. (d) Length of the Fermi arc $l(\phi)$ as a function of temperature. The green arrows mark $T_c$ and $T_p$, and the dashed line is a guide to the eye. The inset shows Lorentzian fit of the angle-dependent spectral function $A(\phi,0)$ on the Fermi surface. (e) Intensity plot of the spectral function $A(\bm{k},0)$ at zero energy in the first Brillouin zone for different temperatures, showing gradual development of the Fermi arc.}
 	\label{fig5}
 \end{figure}

To further compare with experiment \cite{Chen2022}, Fig.~\ref{fig5}(a) plots the energy-momentum dependent spectral function $A(k_y,E)$ at fixed $k_x=-3.047$, which allows us to extract the energy of the maxima for each $k_y$. The resulting dispersions are shown in Fig.~\ref{fig5}(b) for $T=0.098$, 0.07, 0.022. We see that the dispersion exhibits back bending even for $T=0.07>T_c$ but almost recovers the normal state one for $T=0.098>T_p$. The vector $k_G$ where the bending occurs is the same as the Fermi vector $k_{\rm F}=\pm 0.4712$ (the gray vertical line), which differs from the prediction based on density wave or magnetic order pictures. The extracted dispersion also manifests anisotropy due to phase fluctuations. In Fig.~\ref{fig5}(c), the dispersion near $k_F$ (the gray vertical line) shows an angle-dependent gap at $T=0.034$, but a clear node-antinode dichotomy at $T=0.066$, with the near-node dispersion crossing the Fermi energy and the near-antinode dispersion exhibiting a gap and back bending, as reported previously in underdoped experiments \cite{Kanigel2008}. 
 
The effect of the phase correlation is also reflected in the topology of the Fermi surface. As shown in the inset of Fig.~\ref{fig5}(d), the angle-dependent spectral function $A(\phi,0)$ is gradually suppressed away from the nodal point with lowering temperature. This leads to a variation of the Fermi arc \cite{Harrison2007,Alvarez2008,Greco2009}, whose length $l(\phi)$, estimated from the 0.6-maximum width of the spectral peak, is plotted in Fig.~\ref{fig5}(d) as a function of temperature. We see that $l(\phi)$ almost saturates below $T_c$, increases linearly with temperature in the intermediate region, and reaches a full length (Fermi surface) at high temperatures. This confirms its connection with the phase correlation identified using the phase mutual information. Such a temperature variation of the Fermi arc length has been observed in scanning tunneling spectroscopy (STS) experiment \cite{Lee2009}, implying that the zero arc length reported in the ARPES experiments \cite{Kanigel2006} might originate from the peaks of the artificially symmetrized $A(\phi,\omega)$. To be specific, Fig.~\ref{fig5}(e) maps out the zero-energy spectral function $A(\bm{k},0)$ in the first Brillouin zone, and we see a clear evolution from the Fermi arc at $T=0.058$ to the Fermi surface $T=0.09$. This variation indicates that the Bogoliubov quasiparticle appears at different temperatures in different regions of the Fermi surfaces. The arc is more broadened close to the node, consistent with previous experiment \cite{Reber2012}.

\begin{figure}[ptb]
	\begin{center}
		\includegraphics[width=8.6cm]{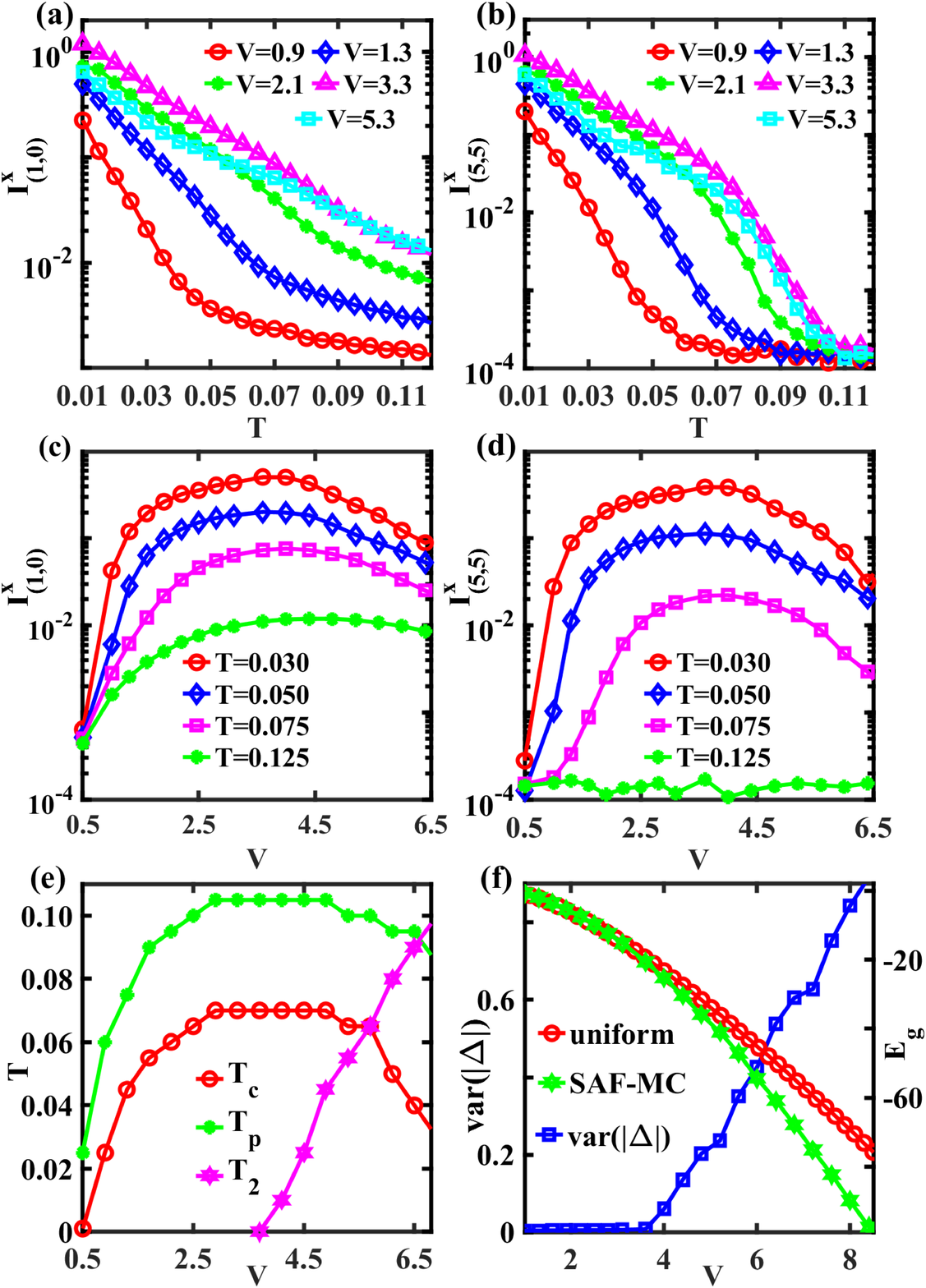}
	\end{center}
	\caption{Comparison of the short- and long-distance phase mutual information with $\bm{R}=(1,0)$ and $(5,5)$ (a)(b) as functions of temperature for different pairing interactions, and (c)(d) as functions of the pairing interactions for different temperatures. (e) The superconducting phase diagram with $T_c$ and $T_p$ determined from the phase mutual information and $T_2$ from the onset of two-peak amplitude distribution.  (f) Comparison of the condensation energy $E_g$ for the uniform mean-field solution and the static auxiliary field Monte Carlo (SAF-MC) solution. Also shown in the variance of the amplitude distribution var$(|\Delta|)$ from the Monte Carlo simulations at $T=0.001$.}
	\label{fig6}
\end{figure}

\subsection{The superconducting phase diagram and a strong-coupling plaquette state}
Having identified the different regions of phase correlations at a fixed $V$, we now turn to their variation with the pairing interaction. As shown in Figs.~\ref{fig6}(a) and ~\ref{fig6}(b), the range of exponential temperature dependence also varies. As $V$ increases, the curves first move to higher temperatures, but then shift somewhat backwards. Such a nonmonotonic variation is better seen in Figs.~\ref{fig6}(c) and \ref{fig6}(d), where  the phase mutual information $I^x_{(1,0)}$ and $I^x_{(5,5)}$ are replotted as a function of $V$ for different temperatures. Both exhibit nonmonotonic behavior with increasing $V$ at low temperatures, indicating that the phase correlations are suppressed when the pairing interaction is getting too large. As we will see, this is closely associated with the two-peak structure of the amplitude distribution in Fig.~\ref{fig1}(d). 

\begin{figure}[ptb]
	\begin{center}
		\includegraphics[width=8.6cm]{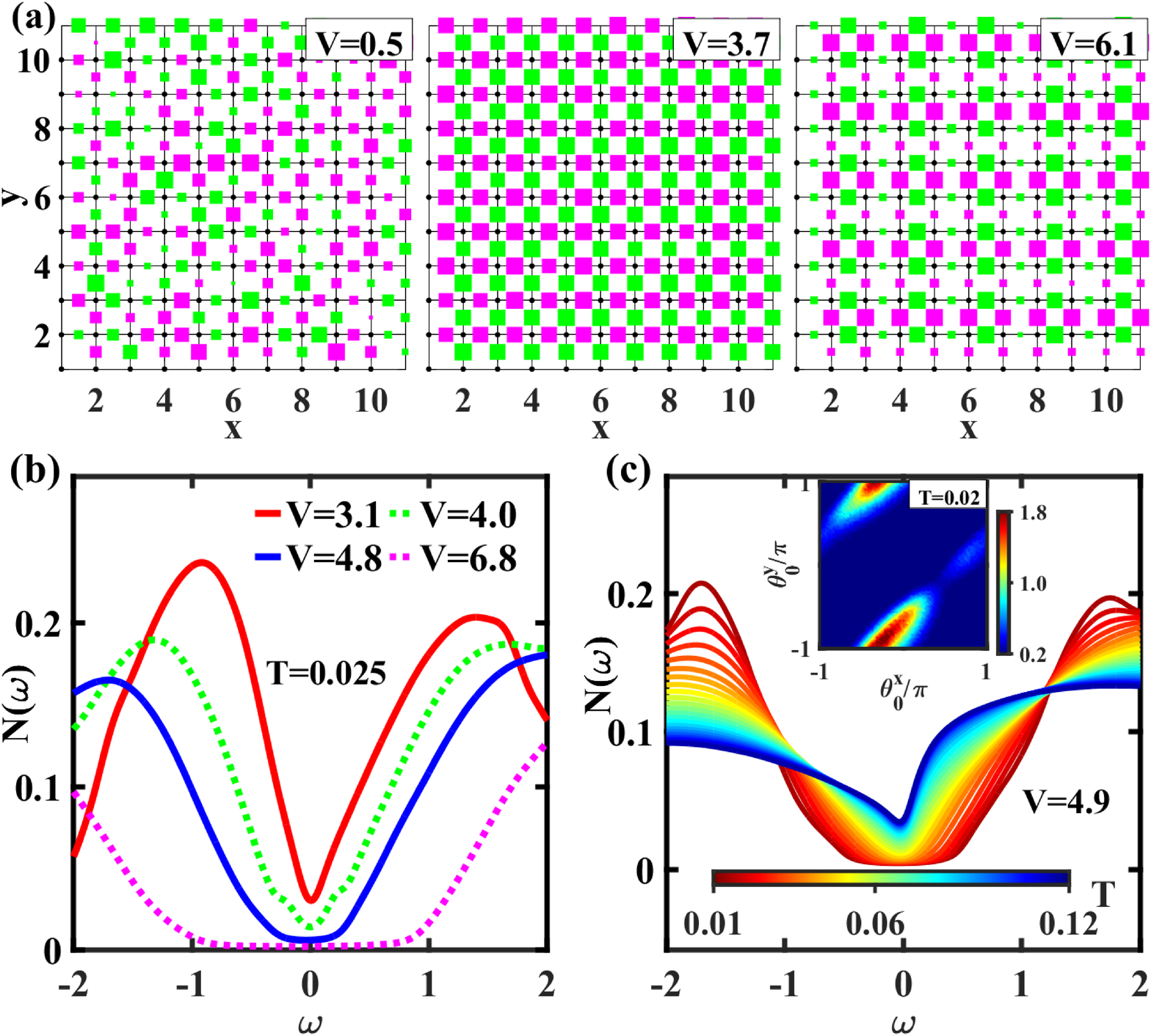}
	\end{center}
	\caption{ (a) Typical configurations of the pairing fields at $T=0.001$ for $V=0.5$, 3.7, 6.1. The square size represents their amplitude and the colors mark the sign of their phase $\theta\in (-\pi,\pi]$. (b) Evolution of the total density of state $N(\omega)$ with pairing interactions at $T=0.025$, showing a crossover from $V$-shape to $U$-shape. (c) Temperature dependence of $N(\omega)$ at $V=4.9$. The inset shows the joint distribution of $\theta^x_{\bm{0}}$ and $\theta^y_{\bm{0}}$ at $T=0.02$, indicating 
			$d$-wave-like bonds for the pairing fields in the plaquette state.}
	\label{fig7}
\end{figure}

Taken together, a superconducting phase diagram can be constructed and shown in Fig.~\ref{fig6}(e), where both $T_c$ and $T_p$ behave nonmonotonically with $V$. 
Also shown is a third temperature scale $T_2$, below which the amplitude distribution has two peaks. $T_2$ only appears for sufficiently large $V$, indicating a strong coupling limit  whose nature will be clarified later. Interestingly, we see that $T_c$ takes its maximum near the critical $V$ of the two-peak distribution and is suppressed as $T_2$ increases. This suggests that the superconductivity is competing with this strong coupling state. 

To clarify this issue, we compare in Fig.~\ref{fig7}(a) typical Monte Carlo configurations of the pairing fields for weak, intermediate, and strong $V$ at $T=0.001$. The size of the square represents the amplitude $|\Delta|$  and the color denotes the sign of the phase $\theta\in(-\pi,\pi]$. For weak $V=0.5$, the distribution on the lattice is random, reflecting that the system is not yet in a phase-coherent region ($T>T_c$). For intermediate $V=3.7$, we find a uniform distribution of the amplitude, while the phase changes sign periodically and exhibits a $d$-wave-like pattern. It is straightforward to identify this state as the uniform $d$-wave superconductivity. For strong $V=6.1$, the amplitude distribution is no longer uniform but exhibits cluster  patterns. We call it a charge-4$e$ $d$-wave plaquette state since it is formed out of local plaquettes \cite{Danilov2022a} with four bonds of large $|\Delta|$ in a unit cell surrounded by weak bonds in a $2\times2$ cell. The plaquette has the same sign structure as the $d$-wave superconductivity. The whole state can be regarded as weakly connected charge-4$e$ plaquettes. Clearly, this is not a phase separation and the two-peak feature of the amplitude distribution is a reflection of the special plaquette structure. This state breaks the translational invariance of the pairing fields, but keeps the uniform distribution of the electron densities. It persists to a very large $V=7.5$, beyond which  the bonds become less correlated  as $t/V\rightarrow 0$.

To show that the plaquette state is stable over the uniform superconductivity, we calculate their condensation energies using  
\begin{equation}
E_g=\sum_l|\xi_l|+\sum_{\langle ij\rangle}\frac{2|\Delta_{ij}|^2}{V}-\sum_{l}\Lambda_l,
\end{equation} 
where $l=1,2,\cdots,~N$ and $\xi_l$ is the eigenvalue of the non-interacting Hamiltonian. Figure~\ref{fig6}(f) compares the condensation energies of the mean-field uniform solution and the Monte Carlo solution. For small $V$, we see they are almost equal. But beyond the critical $V$ of the plaquette state, the mean-field uniform solution has higher energy than the Monte Carlo (plaquette) solution. In this region, the variance var($|\Delta|$) of the amplitude distribution grows rapidly with increasing $V$, reflecting an increasing difference between the strong and weak bonds.

The transition to the plaquette state may be detected from the V-shape-to-U-shape change of the density of states as shown in Fig.~\ref{fig7}(b). Figure~\ref{fig7}(c) plots $N(\omega)$ at $V=4.9$ for different temperatures. The plaquette state melts as $N(\omega)$ changes from U-shape to V-shape with increasing temperature. Note that a U-shaped curve is typically ascribed to $s$-wave superconductivity. However, the plaquette state still exhibits $d$-wave-like bonds as shown in the inset of Fig.~\ref{fig7}(c). A similar variation has been observed in STS measurement in twisted trilayer graphene \cite{Kim2022}, where it was argued to originate from two-particle bound states. In our simulations, the four-particle plaquette state is more favored with nearest-neighbour pairing interaction. 

It has been suggested that strong attractive interaction may always lead to phase separation \cite{Nazarenko1996,Emery1990,Cookson1993,Su2004,Shaw2003}. It could be that the pairing interaction for the plaquette state is not yet strong enough. For sufficiently large $V$, we find randomly distributed dimers and plaquettes, possibly because the pairing correlations are suppressed as $t/V$ becomes too small. The plaquette state may be in some sense related to a pair density wave (PDW) \cite{Lee2014,Setty2021a,Setty2022}. But our derived configuration is special. It does not induce any charge-density wave and may only be produced by a complicated combination of uniform superconductivity and bidirectional PDW states of the wave vector $(0,\pi)$ and $(\pi,0)$. It may thus be better viewed as a different strong-coupling limit of the $d$-wave superconductivity.

\section{Discussion and conclusions} 
We have applied the static auxiliary field Monte Carlo method to study phase correlations of the superconducting pairing fields. We can reproduce the weak-coupling BCS solution of the mean-field theory and identify a region above $T_c$ by separation between short- and long-distance phase correlations for moderate and strong pairing interactions. This phase fluctuating  region above the uniform $d$-wave superconductivity has a number of spectroscopic features including the anisotropy of the angle-resolved gap opening, scattering rate, and specific heat coefficient, as well as gradual development of the Fermi arc. These provide a potential explanation of the experimental observations in overdoped cuprates, and suggest that angular or momentum dependence of the gap opening temperature may be a general feature of phase fluctuations. For sufficiently strong pairing interaction, our simulation reveals a competing charge-4$e$ plaquette state with $d$-wave-like bonds and a U-shaped density of states, which may be useful for understand the pairing in  strong coupling limit.  The superconducting transition temperature seems maximal near the critical pairing interaction of the plaquette state, raising an interesting question concerning their relationship.

While our method captures the superconducting fluctuations at finite temperatures beyond the uniform mean-field theory, it ignores the imaginary-time dependence of the pairing fields and therefore cannot apply at very low temperatures where quantum fluctuations become important. By integrating out the fermions, we only focus on the superconducting properties where nearest-neighbor pairing plays a dominant role. It should be mentioned that we begin the calculations with an attractive spin-singlet pairing interaction. Ignoring other possible instabilities, this effective pairing interaction covers a variety of microscopic pairing mechanisms, including the nearest-neighbor antiferromagnetic spin interaction, the nearest-neighbor attractive charge density interaction \cite{Chen2021f,Plonka2015,Jiang2021a}, and the spin fluctuation mechanism in momentum space \cite{Monthoux2007}. While these mechanisms may be supported by different  experiments \cite{OMahony2022,Chen2021f,Wang2022a}, they exhibit similar superconducting properties as revealed in our calculations.
	
It may be useful to compare our results of the uniform superconductivity with the XY model which is believed to describe the physics of two-dimensional superconductivity \cite{Eckl2002,Han2010,Paramekanti2000}. For this purpose, we have to first define the superconducting order parameter on the lattice sites, namely $\Delta_i=\frac14(\Delta_{i,i+x}+\Delta_{i,i-x}-\Delta_{i,i+y}-\Delta_{i,i-y})$, where $\Delta_{i,i\pm x}$ and $\Delta_{i,i\pm y}$ are the pairing fields on the four bonds connected to site $i$. The number of vortices can then be calculated using $\Delta_i$ following the standard definition \cite{Zhong2011} and found to be nearly zero below $T_c$, grow rapidly between $T_c$ and $T_p$ and slowly above $T_p$, and eventually saturate above $T_l$. The rapid increase above $T_c$ is in good correspondence with that predicted for the BKT transition due to the unbinding of vortices and antivortices, indicating that our $T_c$ is exactly the BKT transition temperature. The power law decay of the phase mutual information indicates a quasi-long range order that does not break U(1) symmetry conforming to the well-known Mermin-Wagner theorem \cite{Mermin1968,Hohenberg1967}. Our identification of three temperature scales and four distinct regions may offer some insight into the triple transition in resistance experment \cite{Rourke2011}, where normal metal, pseudogap (incoherent metal), phase fluctuation, and superconductivity are separated. A similar scenario may also be related to the transition between superconductivity and normal metal, where disorder or magnetic field may broaden the transition and lead to one or two intermediate regions \cite{Spivak2008,Kapitulnik2019}.

Superconducting phase fluctuations also play an important role in other superconductors, such as Fe-based superconductors \cite{Xu2021a,Faeth2021,Kang2020,Grinenko2021a} and disordered conventional superconductors \cite{Mondal2011,Dubouchet2019,Bastiaans2021,Chen2018,Bouadim2011,Ghosal1998,Cui2008}. Our method may also provide useful insight into the interplay between phase fluctuations and other important effects such as disorder, multiband, and time reversal symmetry breaking in these systems. 

This work was supported by the National Natural Science Foundation of China (NSFC Grants No. 11974397, No. 12174429, and No. 12204075), the National Key Research and Development Program of China (Grant No. 2022YFA1402203), and the Strategic Priority Research Program of the Chinese Academy of Sciences (Grant No. XDB33010100).

\appendix
\section{ORIGIN OF THE PAIRING INTERACTION}
We show that the spin singlet pairing interaction may be derived from different microscopic models. To see this, we first define the spin-singlet and spin-triplet pairing operators in real space:
\begin{equation}
\begin{split}
\psi_{ij}^{\rm S}&=\frac{1}{\sqrt{2}}\sum_{ \alpha,\beta}c_{i\alpha}(-i\sigma_y)_{\alpha\beta}c_{j\beta},\\
\bm{\psi}^{\rm T}_{ij}&=\frac{1}{\sqrt{2}}\sum_{ \alpha,\beta}c_{i\alpha}(-i\sigma_y \bm{\sigma})_{\alpha\beta}c_{j\beta},
\end{split}
\end{equation}
which satisfy $\psi_{ji}^{\rm S}=\psi_{ij}^{\rm S}$ and $\bm{\psi}_{ji}^{\rm T}=-\bm{\psi}_{ij}^{\rm T}$.

For nearest-neighbor attractive charge density interaction, we have
\begin{eqnarray}
V_{\rm int}&=&-V^c\sum_{\langle ij\rangle}n_in_j\nonumber\\
&=&-V^c\sum_{\langle ij\rangle}\left[\left(\psi_{ij}^{\rm S}\right)^{\dagger}\psi_{ij}^{\rm S}+\left(\bm{\psi}_{ij}^{\rm T}\right)^{\dagger}\bm{\psi}_{ij}^{\rm T}\right].
\end{eqnarray}
Since this model favors $d$-wave superconductivity \cite{Plonka2015,Jiang2022}, we may discard the spin-triplet part and obtain the singlet pairing interaction in Eq. (\ref{model}) with $V=V^c$.

For nearest-neighbor antiferromagnetic spin density interaction, we have
\begin{equation}
\begin{split}
V_{\rm int}&=\sum_{\langle ij\rangle }J\bm{s}_i\cdot\bm{s}_j\\
&=-\frac{J}{4}\sum_{\langle ij\rangle}\left[3\left(\psi_{ij}^{\rm S}\right)^{\dagger}\psi_{ij}^{\rm S}-\left(\bm{\psi}_{ij}^{\rm T}\right)^{\dagger}\bm{\psi}_{ij}^{\rm T}\right].
\end{split}
\end{equation}
Again, we may exclude the spin-triplet pairing and obtain the singlet pairing interaction in Eq. (\ref{model}) with $V=\frac{3}{4}J$.

The spin fluctuation interaction in momentum space may be transformed into real space, yielding
\begin{eqnarray}
V_{\rm int}=\sum_{\bm{q}} V(\bm{q})\bm{s}_{\bm{q}}\cdot \bm{s}_{-\bm{q}}=\sum_{ij} V(\bm{r}_i-\bm{r}_j) \bm{s}_i\cdot \bm{s}_j,
\end{eqnarray}
where $\bm{s}_{\bm{q}}=\sum_{ \bm{k}}\sum_{ \alpha,\beta}c_{\bm{k}+\bm{q}\alpha}^{\dagger}\frac{\bm{\sigma}_{\alpha,\beta}}{2}c_{\bm{k}\beta}$ and $V(\bm{r}_i)=\sum_{ \bm{q}}V(\bm{q})e^{i\bm{q}\bm{r}_i}$. For some typical phenomenological form of $V(\bm{q})$ \cite{Carbotte1994}, the deduced interaction in real space is dominated by onsite repulsion interaction and nearest-neighbor antiferromagnetic spin density interaction. Excluding the onsite pairing and considering only the nearest-neighbor pairing yield the singlet pairing interaction in Eq. (\ref{model}).

\section{THE EFFECTIVE ACTION}

The action of the Hamiltonian (1) is
\begin{equation}
\begin{split}
S[\bar{c},c]&=\int_{0}^{\beta}d\tau\left[\sum_{il\sigma}\bar{c}_{i\sigma}(\tau)\left((\partial_{\tau}-\mu)\delta_{il}-t_{il}\right)c_{l\sigma}(\tau)\right.\\
&\left.-V\sum_{\langle ij\rangle}\bar{\psi}_{ij}^{\rm S}(\tau)\psi_{ij}^{\rm S}(\tau)\right].
\end{split}
\end{equation}

To decouple the pairing interaction term, we use the Hubbard-Stratonovich transformation \cite{Coleman2015} by introducing the auxiliary pairing field $\Delta_{ij}$ for each nearest-neighbor pair:
\begin{equation}
	\begin{split}
	-&V\bar{\psi}_{ij}^{\rm S}(\tau)\psi_{ij}^{\rm S}(\tau)\rightarrow \\
	&\sqrt{2}\left(\bar{\Delta}_{ij}\psi_{ij}^S(\tau)+\bar{\psi}_{ij}^S(\tau)\Delta_{ij}\right)+\frac{2\left|\Delta_{ij}\right|^2}{V},
	\end{split}
	\end{equation}
where $\bar{\Delta}_{ij}$ is the complex conjugate of $\Delta_{ij}$. In the static approximation, the pairing fields are assumed to be independent of the imaginary time $\tau$. 

We have thus the new action,
\begin{eqnarray}
	S=\sum_{n}\bar{\psi}(i\omega_n)(-i\omega_n+O)\psi(i\omega_n)+ \frac{2\beta}{V}\sum_{\langle ij\rangle}\left|\Delta_{ij}\right|^2,
	\label{action}
	\end{eqnarray}
where $\left[\bar{\psi}(i\omega_n)\right]_j=\bar{c}_{j\uparrow}(i\omega_n)$, $\left[\bar{\psi}(i\omega_n)\right]_{N+j}=c_{j\downarrow}(-i\omega_n)$ for $j=1,2,\cdots N$, and 
\begin{eqnarray}
O=\left(\begin{array}{cc}
-\mu-T&M\\
M^{*}&\mu+T\\
\end{array}\right),
\end{eqnarray}
in which $T_{ij}=t_{ij}$ is the hopping matrix and $M_{ij}=\Delta_{ij}$ contains the pairing term. Integrating out the fermions yields the final effective action,
\begin{equation}
	S_{\rm eff}=-\sum_{ i}\ln(1+e^{-\beta\Lambda_i})+\frac{2\beta}{V} \sum_{\langle ij\rangle}\left|\Delta_{ij}\right|^2,
	\label{Seff}
	\end{equation}
where $\Lambda_i$ is the eigenvalue of $O$. In deriving this expression, we have used
\begin{equation}
\int \mathcal{D}\bar{\psi}\mathcal{D}\psi e^{-\sum_{n}\bar{\psi}(i\omega_n)(-i\omega_n+O)\psi(i\omega_n)}=\prod_{n}\det (O-i\omega_n),
\end{equation}
and
\begin{equation}
\sum_{n}\ln(\Lambda_i-i\omega_n)e^{i\omega_n0^{+}}=\ln(1+e^{-\beta\Lambda_i}).
\end{equation}

\section{THE MEAN FIELD SOLUTION}
Under mean-field approximation, we define $\Delta_{ij}=-\frac{1}{\sqrt{2}}V\langle \psi_{ij}^s\rangle$ and assume a uniform mean-field solution $\Delta_{ij}^x=\Delta_x$ and $\Delta_{ij}^y=\Delta_y$ with $\Delta_x=-\Delta_y=\Delta$. This leads to a $d$-wave gap in the momentum space, $\Delta_{\bm{k}}=2\Delta\left[\cos(k_x)-\cos(k_y)\right]$, that satisfies
\begin{equation}
\Delta=\frac{V}{N}\sum_{\bm{k}}\frac{\Delta_{\bm{k}}}{2E_{\bm{k}}}\left[1-2f(E_{\bm{k}})\right]\cos(k_x),
\end{equation}
where $f(E_{\bm{k}})$ is the Fermi-Dirac distribution function, $E_{\bm{k}}=\sqrt{|\Delta_{\bm{k}}|^2+\xi_{\bm{k}}^2}$, and $\xi_{\bm{k}}$ is the dispersion of an electron.

\section{MONTE CARLO SIMULATIONS}

The effective action (\ref{Seff}) gives the probabilistic distribution of the $2N$ independent complex variables $\Delta_{ij}$:
\begin{eqnarray}
p(\Delta)=Z^{-1}e^{-S_{\rm eff}},~~Z=\int \mathcal{D}\Delta \mathcal{D}\bar{
	\Delta}e^{-S_{\rm eff}},
\end{eqnarray}
which may be simulated using the Monte Carlo approach with the Metropolis algorithm following the standard procedures:

(a) Assign random initial values to the pairing field, $\Delta_{ij}$, on each bond and calculate the matrix $O$ and its eigenvalues.

(b) Update the pairing field to, $|\Delta_{ij}|'=|\Delta_{ij}|+\eta_axV$ or $\theta'_{ij}=\theta_{ij}+\eta_tx$, with $x$ being a random number distributed uniformly between -1 and 1. We find that $\eta_a=1$ and $\eta_t=\pi$ can give an appropriate acceptance rate. 

(c) Calculate the change in the effective action, $\delta S=S_{\rm eff}(\Delta')-S_{\rm eff}(\Delta)$, and accept the update with probability ${\rm min}\{1,\exp(-\delta S)\}$.

(d) Repeat (b) and (c) for 10000 sweeps for thermalization and then 150000 sweeps for measurement. During the measurement, we take one sample for $2N$ pairing fields after every 10 sweeps to reduce the self-correlation effect in the data. All physical quantities are calculated by averaging over 15000 configurations of the pairing fields.

\section{MUTUAL INFORMATION}

The mutual information between two continuous random variables X and Y is defined as \cite{Cover2006}
\begin{equation}
I(X,Y) = \int \int p(x,y) \log \frac{p(x,y)}{p(x) p(y)} dx dy,
\end{equation}
where $p(x,y)$ is their joint probability distribution function, and $p(x)$ and $p(y)$ are their respective marginal probability distribution functions. We have $p(x) = \int p(x,y) dy$ and $p(y) = \int p(x,y) dx$. In practice, we divide the variable domain into intervals and calculate the mutual information by
\begin{equation}
I(X,Y) \approx \sum_{i,j} p_{i,j} \log \frac{p^{xy}_{i,j}}{p^x_i p^y_j},
\end{equation}
where $p^{xy}_{i,j}$, $p^x_i$, and $p^y_j$ are the probabilities for $X$ and $Y$ taking values in the $i$-th and $j$-th intervals, respectively. 

We have chosen 21 intervals in our calculations of the phase mutual information. The results are found to be qualitatively stable if we change the interval size within an appropriate range, or adopt other methods including the adaptive interval partition \cite{Varanasi1999}, the KSG method based on $k$-th nearest neighbors \cite{Kraskov2004PRE}, and the neural network estimation \cite{Belghazi2018PMLR}.

\section{SPECTRAL CALCULATIONS}
To overcome the finite-size effects for spectral property calculations, we apply the twisted boundary conditions \cite{Li2018PRL}:
\begin{equation}
c_{i\sigma}\rightarrow c_{i\sigma}e^{-i\sigma\bm{\phi}\cdot\bm{r}_i},\ \ \ \bm{\phi}=(\frac{2\pi l_x}{n_xm_x},\frac{2\pi l_y}{n_ym_y})
\end{equation}
with $l_x=0,1,\cdots m_x-1$ and $l_y=0,1,\cdots m_y-1$. Here $m_x$ ($m_y$) denotes the number of sublattices in the $x$ ($y$) direction, and $n_x$ ($n_y$) denote the number of lattice sites in the $x$ ($y$) direction of the original lattice. For a square lattice, we have $m_x=m_y$ and $n_x=n_y$. This corresponds to the following transformation:
\begin{eqnarray}
t_{ij}&\rightarrow& t_{ij}e^{i\bm{\phi}\cdot\bm{r}_l},~~~~~\mu \rightarrow \mu, \nonumber\\
\bar{\Delta}_{ij}&\rightarrow& \bar{\Delta}_{ij}e^{i\bm{\phi}\bm{r}_l},~~~~~
\Delta_{ij}\rightarrow \Delta_{ij}e^{i\bm{\phi}\bm{r}_l},
\end{eqnarray}
where $\bm{r}_l = \bm{r}_j - \bm{r}_i$.

In momentum space, the dispersion changes to, $\epsilon_{\bm{k}}\rightarrow\epsilon_{\bm{k}+\bm{\phi}}=-\sum_{ij} t_{ij} e^{i(\bm{k}+\bm{\phi})\cdot\bm{r}_l}$, which effectively includes $m_xn_x\times m_yn_y$ $\bm{k}$-points for calculating physical properties. The pairing fields are then
\begin{eqnarray}
\Delta_{\bm{k}_1+\bm{\phi},\bm{k}_2+\bm{\phi}}=\frac{1}{N}\sum_{ij}\Delta_{ij}e^{i\bm{\phi}\bm{r}_l-i\bm{k}_2\bm{r}_i+i\bm{k}_1\bm{r}_j}.
\end{eqnarray}
We have in the action,
\begin{equation}
O=\left(\begin{array}{cc}
\epsilon_{\bm{k}+\bm{\phi}}-\mu&M\\
M^{\dagger}&-\epsilon_{\bm{k}+\bm{\phi}}+\mu\\
\end{array}\right),
\end{equation}
where $M_{ij}=\Delta_{\bm{k}_j+\bm{\phi},\bm{k}_i+\bm{\phi}}$. This can be diagonalized for each $\bm{\phi}$ and field configuration. The electron Green's function is then obtained after averaging over all field configurations, which gives the spectral function and the density of states presented in the main text.

\end{document}